\documentclass[11pt]{amsart}

\usepackage[foot]{amsaddr}
\usepackage{cite}
\usepackage{fullpage}
\usepackage{graphicx}
\usepackage{amssymb,amsfonts,amsmath,mathrsfs}
\usepackage[version=3]{mhchem}

\newcommand{\eff}{\mathrm{eff}}
\newcommand{\inv}{\mathrm{inv}}
\newcommand{\normdist}{\mathcal{N}}
\newcommand{\obs}{\mathrm{obs}}
\newcommand{\trans}{\mathsf{T}}
\newcommand{\PPPD}{\mathrm{PP/PD}}
\newcommand{\DPDD}{\mathrm{DP/DD}}

\begin{document}

\title{A parameter-free model discrimination criterion based on steady-state coplanarity}

\author{Heather A. Harrington$^{*1}$, Kenneth L. Ho$^{*2}$, Thomas Thorne$^1$, and Michael P.H. Stumpf$^1$}
\address{$^*$ These authors contributed equally}
\address{$^1$ Division of Molecular Biosciences, Imperial College London, Wolfson Building, London, SW7 2AZ, UK}
\address{$^2$ Courant Institute of Mathematical Sciences and Program in Computational Biology, New York University, 251 Mercer Street, New York, NY 10012, USA}%
\email{heather.harrington06@imperial.ac.uk, ho@courant.nyu.edu} 
\email{thomas.thorne04@imperial.ac.uk, m.stumpf@imperial.ac.uk}

\maketitle

\begin{abstract}
 We describe a novel procedure for deciding when a mass-action model is incompatible with observed steady-state data that does not require any parameter estimation. Thus, we avoid the difficulties of nonlinear optimization typically associated with methods based on parameter fitting. Instead, we borrow ideas from algebraic geometry to construct a transformation of the model variables such that any set of steady states of the model under that transformation lies on a common plane, irrespective of the values of the model parameters. Model rejection can then be performed by assessing the degree to which the transformed data deviate from coplanarity. We demonstrate our method by applying it to models of multisite phosphorylation and cell death signaling. Our framework offers parameter-free perspective on the statistical model selection problem, which can complement conventional statistical methods in certain classes of problems where inference has to be based on steady-state data and the model structures allow for suitable algebraic relationships among the steady state solutions. 
\end{abstract}

keywords: chemical reaction networks, Gr\"{o}bner bases, mass-action kinetics, singular values, ordinary differential equations, algebraic statistics.




many branches of science and engineering, one is often interested in the problem of model selection: given observed data and a set of candidate models for the process generating that data, which is the most appropriate model for that process? Such a situation commonly arises when the inner workings of a process are not completely understood, so that multiple models are consistent with the current state of knowledge. For mechanistic models, e.g., ordinary differential equation (ODE) or stochastic dynamical models, most selection techniques involve parameter estimation, which typically requires some form of optimization, exploration of the parameter space, or formal inference procedure \cite{toni:2009:j-r-soc-interface,vyshemirsky:2008:bioinformatics}. For sufficiently complicated models, however, this task can become infeasible, owing to the nonlinearity and multi-modality of the objective function (which penalizes any differences between the data and the model predictions), as well as the high dimensionality of the parameter space \cite{ashyraliyev:2009:febs-j}.

\begin{figure}
 \includegraphics{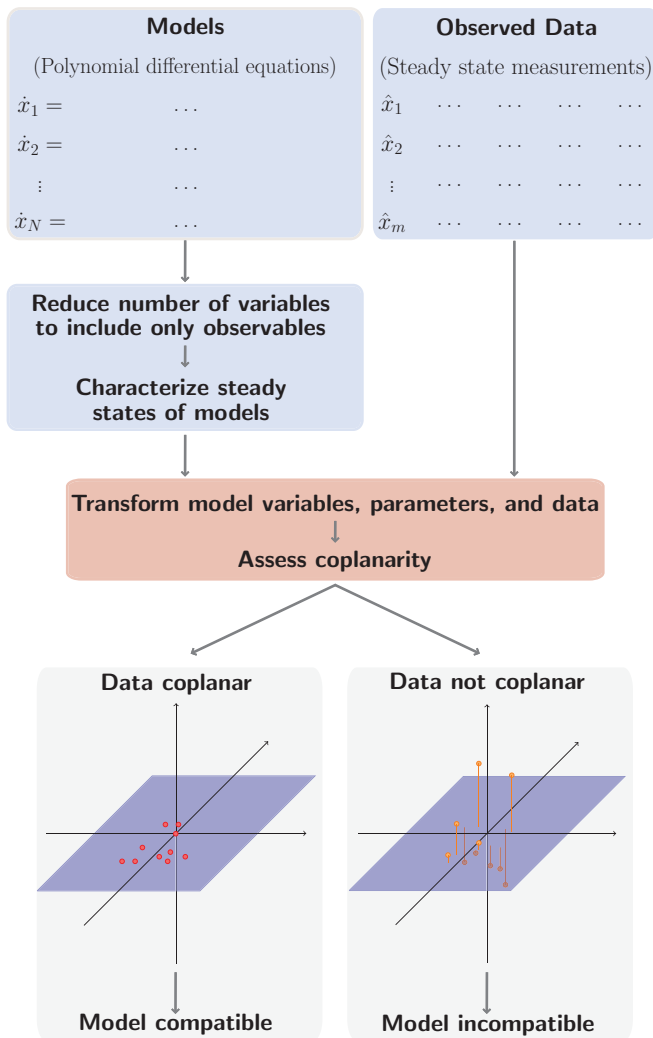}
 \caption{Parameter-free method for model discrimination (see text for details).}
 \label{fig:method}
\end{figure}

Here, we present a framework for the discrimination of mass-action ODE models (and suitable generalizations thereof) that does not require or rely upon such estimated parameters. Our method (Fig.\ \ref{fig:method}) operates on steady-state data and combines techniques from algebraic geometry, linear algebra, and statistics to determine when a given model is incompatible with the data under {\em all} choices of the model parameters. The core idea is to use the model equations to construct a transformation of the original variables such that any set of steady states of the model under that transformation possesses a simple geometric structure, irrespective of parameter values. In this case, we insist that the transformed steady states lie on a plane, which we detect numerically; if the observed data are not coplanar under the transformation induced by a given model, then we can confidently reject that model.

The idea of transformation to coplanarity is not new, but previous efforts were limited, in part, by its systematic detection and quantification. For example, in \cite{manrai:2008:biophys-j}, it was necessary to first manually reduce the dimension of the transformed space to three so that coplanarity could be assessed visually. Other related research using similar methods include \cite{gunawardena:2007:biophys-j,ho:2010:plos-comput-biol,martinez-forero:2010:plos-one}. The current work extends this by devising a numerical scheme for quantifying the deviation from coplanarity that generalizes to higher dimensions and allows for statistical interpretation. Thus, we provide a richer and more powerful framework for the application of this basic technique. Chemical reaction network theory (CRNT) \cite{feinberg:1987:chem-eng-sci,feinberg:1988:chem-eng-sci} and stoichiometric network analysis \cite{clarke:1988:cell-biophys} likewise embrace a parameter-free philosophy and can also be exploited for model selection \cite{conradi:2005:iee-proc-syst-biol,ellison:2000a:j-m-catal-a-chemical,ellison:2000b:j-m-catal-a-chemical}.

It is worth noting that our method provides a necessary but (generally) not sufficient condition for model compatibility: a model that is compatible with the data must provide a transformation to coplanarity, but a model that achieves coplanarity is not necessarily compatible, due to additional degrees of freedom introduced in the transformation process. This is in contrast to traditional approaches based on parameter fitting, which provide a sufficient but not necessary condition since local extrema in the cost function surface may prevent a suitable fit. These two approaches are therefore complementary and can be used together for improved model selection.

The remainder of this paper is organized as follows. First, we introduce the concept of steady-state invariants \cite{manrai:2008:biophys-j,gunawardena:2007:biophys-j}, polynomials that vanish at steady state and which depend only on experimentally accessible variables. Then we illustrate how to use steady-state invariants to deduce coplanarity requirements for model compatibility and how to detect such coplanarity numerically; we also discuss invariants in the context of standard parameter fitting techniques. Next, we apply our method to models of multisite phosphorylation and cell death signaling. Finally, we end with some generalizations and concluding remarks.

\section{Steady-State Invariants}
Consider a chemical reaction network model
\begin{align}
 \sum_{j = 1}^{N} s_{ij} X_{j} \cee{->[k_{i}]} \sum_{j = 1}^{N} s_{ij}' X_{j}, \quad i = 1, \dots, R
\end{align}
in the species $X_{1}, \dots, X_{N}$, where $s_{ij}$ and $s_{ij}'$ are the stoichiometric coefficients of $X_{j}$ in the reactant and product sets, respectively, of reaction $i$, with rate constant $k_{i}$. Under mass-action kinetics, the model has dynamics
\begin{align}
 \dot{x}_{i} = \sum_{j = 1}^{R} k_{j} \left( s_{ji}' - s_{ji} \right) \prod_{k = 1}^{N} x_{k}^{s_{jk}}, \quad i = 1, \dots, N,
\end{align}
where $x_{i}$ is the concentration of species $X_{i}$ (throughout, we follow the convention that lowercase letters denote the concentrations of the corresponding species indicated in uppercase). These equations provide a quantitative description of the model and can, in principle, be used to test its validity by assessing the degree to which they are satisfied by observed data. Unfortunately, in practice, the required variables are rarely all available. In particular, the velocities $\dot{\boldsymbol{x}} = (\dot{x}_{1}, \dots, \dot{x}_{N})$ can be difficult to measure, so we can often consider only the steady state $\dot{\boldsymbol{x}} = \boldsymbol{0}$, as we will do here. Furthermore, certain species may be experimentally inaccessible due to technological limitations; we eliminate these variables from the equations if possible.

For simple models, this elimination can be done by hand, but a more systematic approach is required in general. One such approach is to use Gr\"{o}bner bases \cite{cox:1997:springer}, a central tool in computational algebraic geometry that provides a generalization of Gaussian elimination for multivariate polynomial systems. Here, we follow the general procedure of Manrai and Gunawardena \cite{manrai:2008:biophys-j}. Let $\mathbb{Q} [\boldsymbol{a}]$ be the polynomial ring consisting of all polynomials in the parameters $\boldsymbol{a} = (k_{1}, \dots, k_{R})$ with coefficients from the rational numbers $\mathbb{Q}$, and let $\mathbb{K}$ be its fraction field, comprising all elements of the form $f/g$, where $f, g \in \mathbb{Q} [\boldsymbol{a}]$. Clearly, each $\dot{x}_{i} \in \mathbb{K} [\boldsymbol{x}]$, the ring of all polynomials in $\boldsymbol{x} = (x_{1}, \dots, x_{N})$ with coefficients in $\mathbb{K}$. Note that the parameters $\boldsymbol{a}$ have been absorbed into the coefficient field $\mathbb{K}$; thus, by performing all operations over $\mathbb{K}$, we can treat $\boldsymbol{a}$ symbolically, i.e., without specifying any particular parameter values.

To characterize the steady state $\dot{\boldsymbol{x}} = \boldsymbol{0}$, we construct the ideal $J = \left< \dot{\boldsymbol{x}} \right>$ generated by $\dot{\boldsymbol{x}}$, consisting of all polynomials $\sum_{i = 1}^{N} f_{i} \dot{x}_{i}$, where each $f_{i} \in \mathbb{K} [\boldsymbol{x}]$. Clearly, $J$ contains all elements of $\mathbb{K} [\boldsymbol{x}]$ that vanish at steady state. To obtain only those elements of $J$ that do not depend on the variables $x_{1}, \dots, x_{i}$, we consider the $i$th elimination ideal $J_{i} = J \cap \mathbb{K} [\boldsymbol{x}_{\obs}]$, where $\boldsymbol{x}_{\obs} = (x_{i + 1}, \dots, x_{N})$ denotes the ``observable'' variables. Here, it is useful to introduce Gr\"{o}bner bases, which are special sets of generators with the so-called elimination property that if $\boldsymbol{g} = (g_{1}, \dots, g_{M})$ is a Gr\"{o}bner basis for $J$ under the lexicographic ordering $x_{1} > \cdots > x_{N}$, then $J_{i} = \left< \boldsymbol{g}_{\obs} \right>$, where $\boldsymbol{g}_{\obs} = \boldsymbol{g} \cap \mathbb{K} [\boldsymbol{x}_{\obs}]$ are precisely those elements of $\boldsymbol{g}$ containing only the variables $\boldsymbol{x}_{\obs}$. The polynomials $\boldsymbol{g}_{\obs}$ generate all elements of $\mathbb{K} [\boldsymbol{x}_{\obs}]$ that vanish at steady state and so characterize the projection of the steady state onto the variables $\boldsymbol{x}_{\obs}$.

Procedurally, we compute a reduced Gr\"{o}bner basis $\boldsymbol{g}$ of $J$ with respect to a suitable lexicograhic ordering using standard algorithms, then obtain $\boldsymbol{g}_{\obs}$ by subselection. For numerical convenience we further rescale each polynomial in $\boldsymbol{g}_{\obs}$ so that all coefficients belong to $\mathbb{Q} [\boldsymbol{a}]$ (i.e., we multiply through by their common denominator). Then the elements of $\boldsymbol{g}_{\obs} = (I_{1}, \dots, I_{N_{\inv}})$ have the form
\begin{align}
 I_{i} \left( \boldsymbol{x}_{\obs}; \boldsymbol{a} \right) = \sum_{j = 1}^{n_{i}} f_{ij} \left( \boldsymbol{a} \right) \prod_{k = 1}^{N_{\obs}} x_{k}^{t_{ijk}}, \quad i = 1, \dots, N_{\inv},
\end{align}
where we have applied the relabeling $\boldsymbol{x}_{\obs} = (x_{1}, \dots, x_{N_{\obs}})$. Clearly, each $I_{i}$ is a polynomial in $\boldsymbol{x}_{\obs}$ that vanishes at steady state; we call such polynomials {\em steady-state invariants} (or sometimes just {\em invariants} for short). 

Note in general that steady-state invariants may fail to exist since $J_{i}$ may be empty. Moreover, invariants and their properties (e.g., degrees) can depend delicately on the choice of monomial ordering. Some manual intervention is therefore often required to obtain useful invariants. We will not treat this important (but subtle) issue here, instead focusing on the analysis of given invariants, however they are obtained. This also has the advantage of separating the computation of invariants from their interpretation, in principle allowing the use of invariants from various theories. Steady-state invariants, if they exist, describe relationships between observable variables that hold at steady state for any given realization of parameter values, regardless of other factors such as initial conditions.

For full details on the computational procedure employed, see the accompanying Sage worksheet, which contains code for all computations performed ({\it Materials and Methods}). For further background on algebraic geometry and Gr\"{o}bner bases (including the potential problems of obtaining them), see \cite{cox:1997:springer}; for other methods of variable elimination, see, e.g., \cite{feliu:2012:siam-j-appl-math}. Similar algebraic ideas have also appeared in the context of phylogenetics \cite{cavender:1987:j-classif,pachter:2007:siam-rev}.

\section{Model Discrimination}
We start with a set of steady-state measurements $\hat{\boldsymbol{x}}_{\obs, i}$ for $i = 1, \dots, m$, and a given model with steady-state invariants $\mathscr{I} = \{ I_{1}, \dots, I_{N_{\inv}} \}$.

\subsection{Data Coplanarity}
An invariant, $I \in \mathscr{I}$, can be written somewhat simplified as
\begin{align}
 I \left( \boldsymbol{x}_{\obs}; \boldsymbol{a} \right) = \sum_{j = 1}^{n} f_{j} \left( \boldsymbol{a} \right) \prod_{k = 1}^{N_{\obs}} x_{k}^{t_{jk}}.
 \label{eq:invariant}
\end{align}
We first describe a procedure for deciding whether it is possible that the invariant is compatible with the data, i.e.,
\begin{align}
 I \left( \hat{\boldsymbol{x}}_{\obs, i}; \boldsymbol{a} \right) = 0, \quad i = 1, \dots, m,
 \label{eq:compatibility}
\end{align}
for some choice of $\boldsymbol{a}$. We therefore rewrite \eqref{eq:invariant} as
\begin{align}
 I \left( \boldsymbol{y}; \boldsymbol{b} \right) = \sum_{j = 1}^{n} b_{j} y_{j},
\end{align}
where $y_{j} = \prod_{k = 1}^{N_{\obs}} x_{k}^{t_{jk}}$ and $b_{j} = f_{j} (\boldsymbol{a})$, with $\boldsymbol{y} = (y_{1}, \dots, y_{n})$ and $\boldsymbol{b} = (b_{1}, \dots, b_{n})$. Let $\varphi$ be the map taking $\boldsymbol{x}_{\obs}$ to $\boldsymbol{y}$. Then compatibility implies that the transformed variable $\hat{\boldsymbol{y}} = \varphi (\hat{\boldsymbol{x}}_{\obs})$ corresponding to any observation $\hat{\boldsymbol{x}}_{\obs}$, considered as a point in $\mathbb{R}^{n}$ with coordinates $(\hat{y}_{1}, \dots, \hat{y}_{n})$, lies on the hyperplane defined by the coefficients $\boldsymbol{b}$. In other words, compatibility with the data $\hat{\boldsymbol{x}}_{\obs, i}$ implies that the corresponding transformed data $\hat{\boldsymbol{y}}_{i} = \varphi (\hat{\boldsymbol{x}}_{\obs, i})$ are coplanar.

In general, it is possible that the invariant vanishes trivially, ($\boldsymbol{b} = \boldsymbol{0}$), under some choice of parameters for which coplanarity need no longer hold. To discount this case, we can check, for instance, that the denominator of the corresponding $\boldsymbol{g}_{\obs}$ is never zero. Then $I$ always has at least one nonzero coefficient; hereafter in this section, we assume that the invariant is non-vanishing in this sense.

Let $\boldsymbol{Y} \in \mathbb{R}^{m \times n}$ be the matrix whose rows consist of the $\hat{\boldsymbol{y}}_{i}$. Then the data are coplanar if and only if $\boldsymbol{Y} \boldsymbol{b} = \boldsymbol{0}$ for some nontrivial column vector $\boldsymbol{b} \neq \boldsymbol{0}$. Such a vector, by definition, resides in the null space of $\boldsymbol{Y}$, which can be found using the singular value decomposition $\boldsymbol{Y} = \boldsymbol{U} \boldsymbol{\Sigma} \boldsymbol{V}^{\trans}$, where the diagonal elements of $\boldsymbol{\Sigma}$ give the singular values $\sigma_{i} \geq 0$ encoding the ``stretch'' of each basis vector in $\boldsymbol{V}$. In particular, the smallest singular value $\sigma_{\min}$ bounds the norm $\| \boldsymbol{Y} \boldsymbol{b} \|$ for any $\boldsymbol{b} \neq \boldsymbol{0}$ via
\begin{align}
 \sigma_{\min} = \min_{\left\| \boldsymbol{b} \right\| = 1} \left\| \boldsymbol{Y} \boldsymbol{b} \right\|,
\end{align}
so if $\sigma_{\min} > 0$, then the data cannot be coplanar \cite{belsley:2004:wiley}. More generally, $\sigma_{\min}$ gives the least squares deviation of the data from coplanarity under the scaling constraint $\| \boldsymbol{b} \| = 1$. This quantity depends only on the data and is therefore parameter-free.

Note that this applies for {\em any} choice of $\boldsymbol{b}$, regardless of whether it can be realized by the original parameters $\boldsymbol{a}$. In this sense, the condition of small $\sigma_{\min}$ provides a necessary but not sufficient criterion for model compatibility. The additional degrees of freedom introduced by neglecting the functional forms $f_{j}$ effectively linearizes the compatibility condition \eqref{eq:compatibility}, allowing for a simple, direct solution.

To account for the presence of noise, suppose that we know each component $\hat{x}_{k}$ of a measurement $\hat{\boldsymbol{x}}_{\obs}$ only up to an error $\Delta \hat{x}_{k}$, with
\begin{align}
 \Delta \hat{x}_{k} = \epsilon \hat{x}_{k} Z, \quad k = 1, \dots, N_{\obs},
\end{align}
where $Z \sim \normdist (0, 1)$ is a standard normal random variable. We imagine that the noise parameter $\epsilon$ is given, for example, by instrument error. Then from the perturbation equation
\begin{align}
 \boldsymbol{y} + \Delta \boldsymbol{y} = \varphi \left( \boldsymbol{x}_{\obs} + \Delta \boldsymbol{x}_{\obs} \right),
\end{align}
we find, expanding to first order, that the error is propagated to the transformed variables as $\Delta \hat{\boldsymbol{y}} = \nabla \varphi (\hat{\boldsymbol{x}}_{\obs}) \Delta \hat{\boldsymbol{x}}_{\obs}$, where $\nabla \varphi$ is the Jacobian of $\varphi$, with elements $(\nabla \varphi)_{ij} = \partial y_{i} / \partial x_{j}$. Therefore,
\begin{align}
 \Delta \hat{y}_{j} = \epsilon_{j} Z, \quad \epsilon_{j} = \epsilon \sum_{k = 1}^{N_{\obs}} \left( \nabla \varphi \right)_{jk} \hat{x}_{k}.
 \label{eq:transform-noise}
\end{align}

We now consider the effect of the $\Delta \hat{\boldsymbol{y}}_{i}$ on $\sigma_{\min}$ under the null hypothesis that the underlying $\hat{\boldsymbol{y}}_{i}$ are coplanar with coefficients $\boldsymbol{b}$ (of unit norm). For this, we study the vector $\boldsymbol{Y} \boldsymbol{b}$, whose entries are perturbed from zero to
\begin{align}
 \sum_{j = 1}^{n} b_{j} \Delta \hat{y}_{j} = \left( \sum_{j = 1}^{n} b_{j} \epsilon_{j} \right) Z
 \label{eq:invariant-noise}
\end{align}
for each transformed datum $\hat{\boldsymbol{y}}$. Since $\| \boldsymbol{b} \| = 1$ by assumption, if we rescale each row of $\boldsymbol{Y}$ by its corresponding effective error
\begin{align}
 \epsilon_{\eff} = \max_{j = 1, \dots, n} \left| \epsilon_{j} \right| \geq \left| \sum_{j = 1}^{n} b_{j} \epsilon_{j} \right|,
\end{align}
thus obtaining $\boldsymbol{Y}'$, then each entry of $\boldsymbol{Y}' \boldsymbol{b}$ has the form $\mu_{i} Z$ with $|\mu_{i}| \leq 1$, for $i = 1, \dots, m$. We hence define the {\em coplanarity error}
\begin{align}
 \Delta = \sigma_{\min} \left( \boldsymbol{Y}' \right) \leq \left\| \boldsymbol{Y}' \boldsymbol{b} \right\|,
\end{align}
which, from the discussion above, is bounded by the length of a normal random vector with variances $\mu_{i}^{2} \leq 1$, whose distribution function clearly dominates that of the length of a normal random vector with variances $\mu_{i}^{2} = 1$. But this latter quantity simply follows the $\chi$ distribution with $m$ degrees of freedom. In other words,
\begin{align}
 \Pr \left( \Delta \geq x \right) \leq \Pr \left( X \geq x \right), \quad X \sim \chi_{m};
\end{align}
if $p_{\alpha}$ is the upper $\alpha$-percentile for $\chi_{m}$ (e.g., $\alpha = 0.05$), then
\begin{align}
 \Pr \left( \Delta \geq p_{\alpha} \right) \leq \Pr \left( X \geq p_{\alpha} \right) = \alpha,
\end{align}
which gives an approximate criterion for rejecting coplanarity. As the amount of data increases, the approximation improves since $\sigma_{\min} (\boldsymbol{Y}') \to \| \boldsymbol{Y}' \boldsymbol{b} \|$ as $m \to \infty$ by the symmetry of \eqref{eq:transform-noise}.

Depending on the exact situation at hand, it may be appropriate to choose a more conservative significance level $\alpha$ or to invoke additional criteria in order to decide whether a model is acceptable. In the examples below, however, we will see that whether a model can be rejected is often fairly obvious, and in such cases we will simply use the asymptotic arguments based on the $\chi_{m}$ distribution.

\subsection{Invariant Minimization}
Steady-state invariants can also be used in conjunction with standard parameter fitting techniques. The basic approach is to minimize the Frobenius norm of the matrix $\boldsymbol{\theta} \in \mathbb{R}^{N_{\inv} \times m}$, with entries $\boldsymbol{\theta}_{ij} = I_{i} (\hat{\boldsymbol{x}}_{\obs, j}; \boldsymbol{a})$, over the parameters $\boldsymbol{a}$. This readily provides a sufficient condition for model compatibility since any $\boldsymbol{a}$ producing a small norm provides parameters that fit the data by construction. However, the condition is not necessary since suitable parameters may fail to be found even for compatible models due to the intricacies of the objective function. Clearly, prior knowledge of $\boldsymbol{a}$ can be used to guide the optimization away from such difficulties.

Assuming that the model and its parameters are correct, each invariant $I(\hat{\boldsymbol{x}}_{\obs}; \boldsymbol{a}) = 0$ in principle. However, due to noise, $I(\hat{\boldsymbol{x}}_{\obs}; \boldsymbol{a}) = \epsilon (\hat{\boldsymbol{x}}_{\obs}; \boldsymbol{a}) Z$, where
\begin{align}
 \epsilon \left( \hat{\boldsymbol{x}}_{\obs}; \boldsymbol{a} \right) = \epsilon \sum_{j = 1}^{n} f_{j} \left( \boldsymbol{a} \right) \sum_{k = 1}^{N_{\obs}} \left( \nabla \varphi \right)_{jk} \hat{x}_{k}
\end{align}
by \eqref{eq:invariant-noise}. Therefore, if we use $I(\hat{\boldsymbol{x}}_{\obs}; \boldsymbol{a}) / \epsilon (\hat{\boldsymbol{x}}_{\obs}; \boldsymbol{a})$ as the entry of $\boldsymbol{\theta}$ corresponding to invariant $I$ and datum $\hat{\boldsymbol{x}}_{\obs}$, then the {\em invariant error}
\begin{align}
 \theta \left( \boldsymbol{a} \right) = \left\| \boldsymbol{\theta} \left( \boldsymbol{a} \right) \right\|_{F} \sim \chi_{N_{\inv} m}.
\end{align}
This can be used to compute the likelihood $L(\boldsymbol{a}) = \Pr (\theta (\boldsymbol{a}))$ and allows, e.g., various likelihood-based selection schemes \cite{casella:2001:duxbury,schwarz:1978:ann-statist}, assuming that the optimization can be performed. Here, we use the Akaike information criterion (AIC),
\begin{align}
 A = 2R - 2 \log L_{\max},
\end{align}
where $L_{\max} = \max_{\boldsymbol{a}} L(\boldsymbol{a})$, which penalizes model complexity; the preferred model is the one with the minimum AIC \cite{akaike:1974:ieee-trans-automat-control}.

\section{Results}
We apply our methods to two illustrative biological processes for which competing models exist: multisite phosphorylation and cell death signaling.

\subsection{Multisite Phosphorylation}
We focus first on phosphorylation, a key cellular regulatory mechanism that has been the subject of extensive study, both experimentally \cite{cohen:1982:nature,cohen:2000:trends-biochem-sci,seger:1995:faseb-j} and theoretically \cite{manrai:2008:biophys-j,gunawardena:2007:biophys-j,gunawardena:2005:proc-natl-acad-sci-usa,huang:1996:proc-natl-acad-sci-usa,thomson:2009:nature}. Following \cite{manrai:2008:biophys-j}, we consider a two-site system with reactions,
\begin{subequations}
 \begin{align}
  K + S_{u} \cee{<=>[a_{u}][b_{u}]} K S_{u} \cee{->[c_{uv}]} K + S_{v},\\
  F + S_{v} \cee{<=>[\alpha_{v}][\beta_{v}]} F S_{v} \cee{->[\gamma_{vu}]} F + S_{u},
 \end{align}
\end{subequations}
where $u, v \in \{ 0, 1 \}^{2}$ are bit strings of length two, encoding the occupancies of each site ($0$ or $1$ for the absence or presence, respectively, of a phosphate), with $u$ having less bits than $v$; $S_{u}$ is the phosphoform with phosphorylation state $u$; $K$ is a kinase, an enzyme that adds phosphates; and $F$ is a phosphatase, an enzyme that removes phosphates. Each enzyme can be either {\em processive} (P), where more than one phosphate modification may be achieved in a single step, or {\em distributive} (D), where only one modification is allowed before the enzyme dissociates from the substrate ($c_{0011} = 0$ for $K$, $\gamma_{1100}$ for $F$). This mechanistic diversity generates four competing models: PP, PD, DP, and DD; where the first letter designates the mechanism of the kinase, and the second, that of the phosphatase.

As in \cite{manrai:2008:biophys-j}, we consider only the concentrations $\boldsymbol{x}_{\obs} = (s_{00}, s_{01}, s_{10}, s_{11})$ as observable and use the ordering,
\begin{align}
 (ks_{00}, ks_{01}, ks_{10}, fs_{01}, fs_{10}, fs_{11}, k, f, s_{00}, s_{01}, s_{10}, s_{11}),
 \label{eq:phos-order}
\end{align}
with which we are able to eliminate all other variables except $f$ from the dynamics of each model. The remaining Gr\"{o}bner basis polynomials are of the form $p(f, \boldsymbol{x}_{\obs}) = f \cdot q(\boldsymbol{x}_{\obs})$, where $f \neq 0$ unless there is no phosphatase in the system, which we assume not to be the case, so we take only the observable part $q(\boldsymbol{x}_{\obs})$. It is easy to check that the resulting denominators are always of one sign.

Each model has three steady-state invariants. Matched appropriately, the invariants for model PP share the same transformed variables $\boldsymbol{y} = \varphi (\boldsymbol{x}_{\obs})$ as those for PD; the same is true for DP and DD. Thus, in terms of the transformed data, only the kinase mechanism is discriminative. Between PP/PD and DP/DD, two invariants ($I_{1}$ and $I_{2}$) are discriminative in principle, though only one ($I_{2})$ succeeds numerically: for simulated data from the PP/PD models, provided that the noise level is sufficiently low, lack of coplanarity on $I_{2}$ is able to correctly reject the DP/DD models at significance level $\alpha = 0.05$ ($\Delta \sim 10^{5}$ versus $\Delta \sim 1$ for PP/PD at $\epsilon = 10^{-9}$, against a threshold of $p_{\alpha} = 11.2$). The corresponding test using DP/DD data is not successful due to the form of $I_{2}$, which has transformed variables,
\begin{subequations}
 \begin{align}
  \boldsymbol{y}^{\PPPD} &= \left( s_{00} s_{10}, s_{00} s_{11}, s_{01} s_{10}, s_{01} s_{11}, s_{10}^{2}, s_{10} s_{11} \right),\\
  \boldsymbol{y}^{\DPDD} &= \left( s_{00} s_{11}, s_{01} s_{10}, s_{01} s_{11}, s_{10}^{2}, s_{10} s_{11} \right)
 \end{align}
\end{subequations}
for PP/PD and DP/DD, respectively, i.e., $\boldsymbol{y}^{\PPPD}$ has the additional variable $s_{00} s_{10}$ over $\boldsymbol{y}^{\DPDD}$. Therefore, PP/PD models can be made to fit DP/DD data simply by setting the coefficient corresponding to $s_{00} s_{10}$ to zero, which is in fact what we observe. No model is rejected on the basis of data generated from it.

We emphasize that these results are specific to the particular ordering chosen. Indeed, one can make the phosphatase mechanism discriminative instead by reversing the order of the variables $\boldsymbol{x}_{\obs}$ in \eqref{eq:phos-order}. The exhaustive analysis of such orderings is beyond the scope here; rather, we aim to illustrate the potential uses (and usefulness) of this type of approach using concrete examples. 

\begin{figure}
 \includegraphics{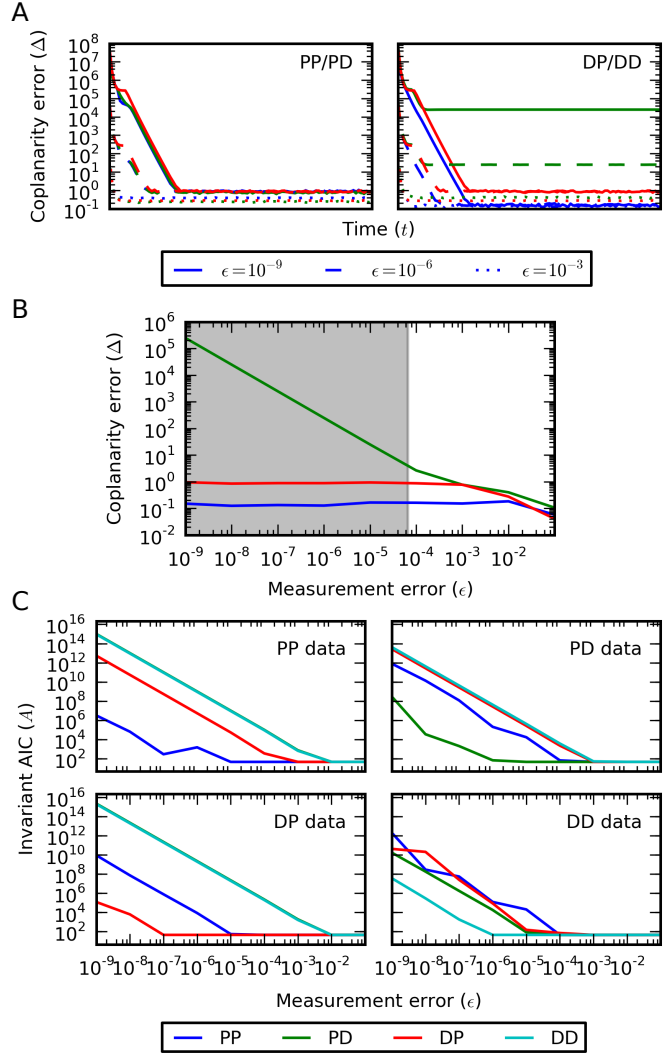}
 \caption{Discrimination of multisite phosphorylation models. ({\it A}) Coplanarity error $\Delta$ of the steady-state invariants of the PP/PD (left) and DP/DD (right) models along time course trajectories simulated from the PP model, corrupted by various levels of noise (lined, $\epsilon = 10^{-9}$; dashed, $\epsilon = 10^{-6}$; dotted, $\epsilon = 10^{-3}$). At each noise level, the errors for three invariants are shown (blue, $I_{1}$; green, $I_{2}$; red, $I_{3}$). ({\it B}) Coplanarity error $\Delta$ of DP/DD invariants on PP data at steady state as a function of the noise level $\epsilon$; invariants colored as in (A). The shaded region indicates the regime over which the DP/DD models can be rejected at significance level $\alpha = 0.05$. ({\it C}) Invariant error AIC $A$ for each model (blue, PP; green, PD; red, DP; cyan, DD) on data generated from the PP (top left), PD (top right), DP (bottom left), and DD (bottom right) models.}
 \label{fig:phos}
\end{figure}

Although the condition of coplanarity is technically valid only at steady state,  there should nevertheless be some convergence over time to coplanarity for any compatible model. We hence compute $\Delta$ for the PP/PD and DP/DD models along time course trajectories simulated from model PP at various levels of $\epsilon$ (Fig.\ \ref{fig:phos}A). For low noise, the results confirm convergence for invariants previously identified as compatible (all $I_{i}$ for PP/PD; $I_{1}$ and $I_{3}$ for DP/DD), with stagnation for incompatible invariants ($I_{2}$ for DP/DD); this does suggest wider applicability of this method, provided that the data are approaching steady state reasonably fast. As the noise increases, however, $\Delta$ decreases inversely proportionally, until the stagnation point hits the basal error level of $\Delta \sim 1$ and we lose all power to reject. Additional simulations estimate the critical noise level at $\epsilon \sim 10^{-4}$ (Fig.\ \ref{fig:phos}B).

To further discriminate between all four models we next turn to invariant minimization. The required optimization involves highly nonlinear functions, so success should be expected only if we have good initial estimates of the model parameters. This is a rather strong demand. In such a case, however, minimization is indeed capable of identifying the correct model from the data so long as $\epsilon \lesssim 10^{-5}$ (Fig.\ \ref{fig:phos}C). These results reinforce our belief that the algebraic approach proposed here naturally complements conventional (i.e. parametric) reverse engineering schemes such as optimization or inference procedures.

\subsection{Cell Death Signaling}
We next apply our methods to receptor-mediated cell death signaling, the so-called extrinsic apoptosis pathway, which plays a prominent role in cancers and other diseases \cite{thompson:1995:science,raff:1998:nature,meier:2000:nature,fulda:2006:oncogene}. Specifically, we consider the assembly of the death-inducing signaling complex (DISC), a multi-protein oligomer formed by the association of FasL, a death ligand, with its cognate receptor Fas \cite{ashkenazi:1998:science,peter:2003:cell-death-differ}.

We investigate two models of DISC formation. The first \cite{lai:2004:math-biosci-eng}, which we call the {\em crosslinking model}, is based on the successive binding of Fas ($R$) to FasL ($L$),
\begin{subequations}
 \begin{align}
  L + R &\cee{<=>[{3 k_{f}}][k_{r}]} C_{1},\\
  C_{1} + R &\cee{<=>[{2 k_{f}}][{2 k_{r}}]} C_{2},\\
  C_{2} + R &\cee{<=>[k_{f}][{3 k_{r}}]} C_{3},
 \end{align}
\end{subequations}
where $C_{i}$ is the complex FasL:Fas$_{i}$. The second \cite{ho:2010:plos-comput-biol}, which we call the {\em cluster model}, posits three forms of Fas (inactive, $X$; active and unstable, $Y$; active and stable, $Z$) and specifies receptor cluster-stabilization events driven by FasL,
\begin{subequations}
 \begin{align}
  X &\cee{<=>[k_{o}][k_{c}]} Y,\\
  Z &\cee{->[k_{u}]} Y,\\
  jY + \left( i - j \right) Z &\cee{->[k_{s}^{\left( i \right)}]} \left( j - k \right) Y + \left( i - j + k \right) Z,\\
  L + jY + \left( i - j \right) Z &\cee{->[k_{l}^{\left( i \right)}]} L + \left( j - k \right) Y + \left( i - j + k \right) Z,
 \end{align}
\end{subequations}
where the last two reactions represent entire families generated by taking $i = 2$ or $3$, with $j = 1, \dots, i$ and $k = 1, \dots, j$. The cluster model is capable of bistability, whereas the crosslinking model exhibits only monostable behavior \cite{ho:2010:plos-comput-biol}.

The two models are structurally very different, and discriminating between them requires some care. Hence, following \cite{ho:2010:plos-comput-biol}, we establish a correspondence between the models by considering the apoptotic signal $\zeta$ transduced by the DISC, defined as $\zeta = c_{1} + 2 c_{2} + 3 c_{3}$ for the crosslinking model and $\zeta = z$ for the cluster model. We assume that $\zeta$ is experimentally accessible; other variables assumed accessible include $\lambda$, the total concentration of FasL ($\lambda = l + c_{1} + c_{2} + c_{3}$ and $\lambda = l$ for the crosslinking and cluster models, respectively), and $\rho$, the total concentration of Fas ($\rho = r + c_{1} + 2 c_{2} + 3 c_{3}$ and $\rho = x + y + z$, respectively). Eliminating all other variables via the orderings $(c_{2}, c_{3}, \lambda, \rho, \zeta)$ and $(y, \lambda, \rho, \zeta)$ for the crosslinking and cluster models (after appropriate variable substitutions) we obtain one non-vanishing steady-state invariant for each model. The dimensions of the transformed spaces are $5$ and $15$ for the crosslinking and cluster models, respectively.

\begin{figure}
 \includegraphics{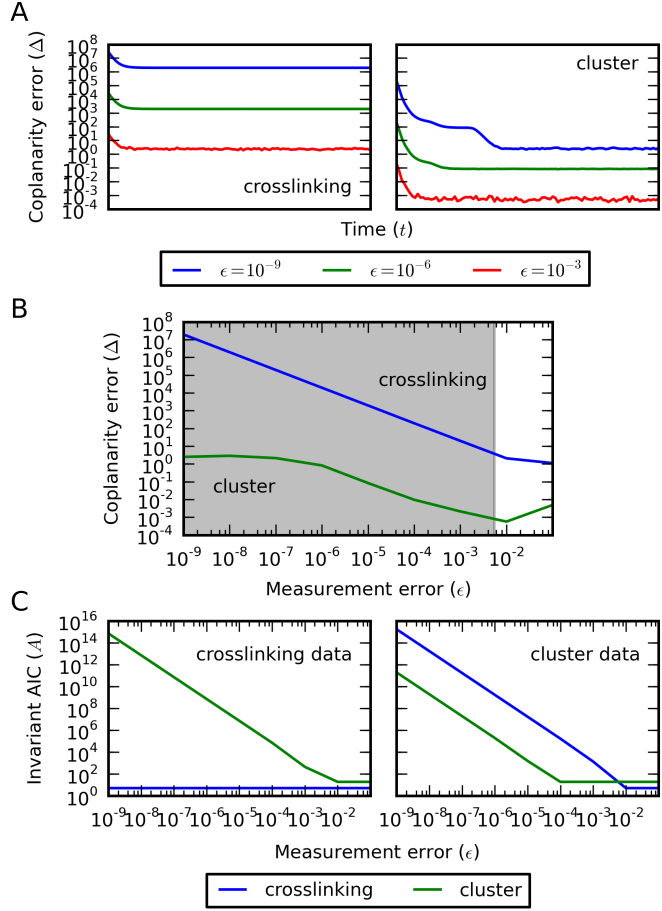}
 \caption{Discrimination of cell death signaling models. ({\it A}) Coplanarity error $\Delta$ of the steady-state invariants of the crosslinking (left) and cluster (right) models along time course trajectories simulated from the cluster model, corrupted by various levels of noise (blue, $\epsilon = 10^{-9}$; green, $\epsilon = 10^{-6}$; red, $\epsilon = 10^{-3}$). ({\it B}) Coplanarity error $\Delta$ of model invariants (blue, crosslinking; green, cluster) on cluster data at steady state as a function of the noise level $\epsilon$. The shaded region indicates the regime over which the crosslinking model can be rejected at significance level $\alpha = 0.05$. ({\it C}) Invariant error AIC $A$ for each model (blue, crosslinking; green, cluster) on data generated from the crosslinking (left) and cluster (right) models.}
 \label{fig:apop}
\end{figure}

As for phosphorylation, we compute the coplanarity error for each invariant on time course data simulated from each model at various noise levels. Although results are inconclusive for data from the crosslinking model, the coplanarity criterion can reject the crosslinking model on the basis of cluster model data at $\alpha = 0.05$, provided that $\epsilon \lesssim 10^{-2}$ (Fig.\ \ref{fig:apop}A and B). The minimization protocol also correctly identifies the model from the data over the same range of noise levels (Fig.\ \ref{fig:apop}C).

\section{Discussion}
In this paper, we have presented a novel model discrimination scheme based on steady-state coplanarity that does not require known or estimated parameter values. Thus we are able to sidestep the parameter inference problem common to many fields including systems biology \cite{ashyraliyev:2009:febs-j,gunawardena:2010:wiley}. Such algebraic methods are not always effective, however: steady-state invariants may not exist, and even when they do, the additional degrees of freedom introduced by effective linearization can cause the method to fail. A promising solution to the problem when invariants cannot be calculated using Gr\"{o}bner bases may be to employ invariants from CRNT \cite{karp:2012:j-theor-biol}. Our results also suggest a somewhat low tolerance for noise, which can restrict its applicability.  Significantly, our method has the unique feature that it can be applied with complete ignorance of parameter values, and is therefore a useful additional tool in the analysis of inverse problems involving dynamical systems.

Rather than competing directly with current model discrimination techniques, we expect that coplanarity will form one end of an entire spectrum of methods, to be used when  no parameter information is available. At the other end lie methods based on parameter estimation (including invariant minimization), which, for dynamical systems, can depend delicately on qualitative and quantitive aspects of the systems under consideration \cite{erguler:2011:mol-biosyst,gutenkunst:2007:plos-comput-biol}. The intermediate regime comprises techniques that can leverage partial knowledge, for instance, constraints on certain parameter values or qualitative features of the dynamics \cite{melykuti:2010:bmc-syst-biol}. Along this spectrum, naturally, the discriminative power increases with the amount of prior information available. In this broader context, coplanarity can be used to efficiently reject candidate models before employing more demanding parameter estimation tools. Thus, it can serve as a preprocessor to thin out the model space. { The real advantages and limitations of any inferential procedure become apparent once their performance can be   evaluated in real-world applications. This is perhaps particularly true for this current approach. Certainly a range of theoretical and computational issues surround algebraic methods which will likely impact their applicability. Here we have found that a pragmatic approach yields some useful insights for small and intermediate-sized problems.}

Finally, we remark that the presented scheme is but the simplest of a potential new class of parameter-free selection methods based on the detection of geometric structure. In this view, transformation to coplanarity is just one of many low-dimensional descriptions of such structure. The existence of low-dimensional representations has recently been predicted in neuronal signaling \cite{barbano:2007:proc-natl-acad-sci-usa}, and can ultimately be attributed to the inherent robustness of biological systems \cite{csete:2002:science,kitano:2004:nat-rev-genet}.


\section{Materials}
 \subsection{Gr\"{o}bner Basis Calculation}
 All reduced Gr\"{o}bner bases are computed over the field $\mathbb{K}$ of rational functions in the parameters $\boldsymbol{a}$ with rational coefficients, under a suitable lexicographic ordering with the observables $\boldsymbol{x}_{\obs}$ located at the end of the variable list, using the computer algebra system {\sc Singular} (http://www.singular.uni-kl.de/) as interfaced through Sage (http://www.sagemath.org/).

 \subsection{Data Generation}
 For each model parameters are drawn independently from a log-normal distribution with median $\mu^{*} = e^{\mu} = 1$ and multiplicative standard deviation $\sigma^{*} = e^{\sigma} = 2$, where $\mu$ and $\sigma$ are the mean and standard deviation, respectively, of the underlying normal distribution. Using these parameters $m = 100$ time course trajectories are computed for each model via integration of the model ODEs over the time interval $0 \leq t \leq 100$; each trajectory is seeded by random initial conditions sampled from a log-normal distribution also with $\mu^{*} = 1$ and $\sigma^{*} = 2$. Integration is performed using the solver LSODA as wrapped in SciPy (http://www.scipy.org/). The data are then corrupted by noise of varying levels from $\epsilon = 10^{-9}$ to $10^{-1}$, for each $\epsilon$, multiplying the nominal data by random log-normal samples with $\mu^{*} = 1$ and $\sigma^{*} = 1 + \epsilon$.

 \subsection{Invariant Minimization}
 Invariant error likelihood maximization is performed in two phases. First, an approximate optimal parameter set is obtained by minimizing the Frobenius norm of the matrix $\boldsymbol{\eta} \in \mathbb{R}^{N_{\inv} \times m}$, where each entry corresponds to an invariant-datum pair as in $\boldsymbol{\theta}$, but with values $I(\hat{\boldsymbol{x}}_{\obs}; \boldsymbol{a}) / M(\hat{\boldsymbol{x}}_{\obs}; \boldsymbol{a})$, where
 \begin{align}
  M \left( \hat{\boldsymbol{x}}_{\obs}; \boldsymbol{a} \right) = \sum_{j = 1}^{n} \left| f_{j} \left( \boldsymbol{a} \right) \prod_{k = 1}^{N_{\obs}} \hat{x}_{k}^{t_{jk}} \right|.
 \end{align}
 This is then taken as an initial parameter estimate to compute $L_{\max}$. All optimizations are performed using L-BFGS-B \cite{zhu:1997:acm-trans-math-softw} through SciPy, with lower and upper bounds of $0.01$ and $100$, respectively, for each variable. The minimization of $\| \boldsymbol{\eta} \|_{F}$ is seeded with initial value $1$ for all variables.

 \subsection{Computational Platform}
 All computations are performed centrally in Sage, making use of its interfaces to various programs. Plots were produced using matplotlib (http://matplotlib.sourceforge.net/). The Sage worksheet for this paper, which contains code for all computations performed, is available at http://www.sagenb.org/home/pub/3462/.


\section{Acknowledgments}
 HAH and MPHS gratefully acknowledge the Leverhulme Trust. KLH was funded in part by National Science Foundation grant DGE-0333389. TT and MPHS also acknowledge funding from the Biotechnology and Biological Research Council; MPHS is a Royal Society Wolfson Research Merit award holder. We thank Carsten Wiuf, Elisenda Feliu, and Sarah Filippi for their comments on the manuscript.





\end{document}